\documentclass[a4paper,fleqn,usenatbib]{mnras}

\usepackage{newtxtext,newtxmath}

\usepackage[T1]{fontenc}
\usepackage{ae,aecompl}

\usepackage{graphicx}	
\usepackage{amsmath}	
\usepackage{amssymb}	
\usepackage{chemformula}

\newcommand{\Htwo}{\ch{H2}}
\newcommand{\HI}{H{\sc i}}
\newcommand{\HII}{H{\sc i}\,21cm}
\newcommand\casA{Cassiopeia~A}
\newcommand \kmps{km~$\text{s}^{-1}$ }
\newcommand\CO{\ch{CO}}
\newcommand{\HCO}{\ch{H2CO}}
\newcommand{\NH}{\rm{N}_{\Htwo}}

\title[Mapping CRRLs towards Cas~A]{A High Resolution Study of Carbon Radio Recombination Lines towards Cassiopeia~A}

\author[A. Chowdhury \& J.N. Chengalur]{
Aditya Chowdhury  $^{1}$\thanks{E-mail: chowdhury@ncra.tifr.res.in} \&
Jayaram N. Chengalur$^{1}$\thanks{E-mail: chengalur@ncra.tifr.res.in}
\\
$^{1}$National Centre for Radio Astrophysics, Tata Institute of Fundamental Research (NCRA-TIFR), Pune, India\\
}

\pubyear{2018}

\begin{document}
\label{firstpage}
\pagerange{\pageref{firstpage}--\pageref{lastpage}}
\maketitle

\begin{abstract}
Carbon Radio Recombination Lines trace the interface between molecular and atomic gas. We present GMRT observations of Carbon Radio Recombination Lines (CRRL), C$\alpha$244-C$\alpha$250 towards Cassiopeia A. We use a novel technique of stacking the emission lines in the visibility domain to obtain, for the first time, sub-pc resolution optical depth maps of these CRRLs. The emission shows a wide range of spatial and velocity structures, some of which are unresolved within our synthesis beam of 0.29 pc and velocity channel of 0.55 \kmps. {These variations in the emission optical depth and line width} are indicative of inhomogeneity and fragmentation in the intervening Perseus Arm gas. We compare the distribution of the CRRL emission with that of diffuse and dense molecular gas using existing CO and \HCO \ studies. We find that the \CO \ emission in the -47 \kmps Perseus Arm component is primarily concentrated along an elongated structure detected in our CRRL maps, to the south of which lies the high-density molecular clumps traced by \HCO. This spatial distribution of CRRL and molecular tracers is similar to what one would observe for a Photo Dissociation Region (PDR). In the other Perseus Arm component centered at -37 \kmps, there is evidence for { high column density} ($\NH \sim10^{22}$cm$^{-2}$) molecular clumps embedded in { diffuse \CO \ as} well as CRRL emission towards the center of Cassiopeia A. { We propose that the CRRL emissions coincident with molecular tracers originates from the line of sight integrated component of the \ch{C+} envelope of the molecular gas.}
\end{abstract}




\section{Introduction}

The interstellar medium shows a rich diversity of physical conditions ranging from a very hot state where atomic species are found in the plasma state to an extremely cold state where complex organic molecules are formed. These range of conditions are intricately tied to star formation wherein stars are formed in the densest regions of molecular clouds and energy input from stars leads to heating $\&$ ionization of the gas. Observations suggest that these molecular clouds are embedded in regions where atomic hydrogen is abundant [for example, \citet{Fukui09} provides evidence for gravitationally bound \HI \ around Giant Molecular Clouds in the Large Magellanic Clouds]. This association between molecular and atomic gas can be explained via the \HI-to-\Htwo \ conversion process in regions of atomic clouds which are self-shielded from radiation that may cause photo-dissociation of the molecules \citep[e.g.][]{Gould63,Wakelam17}. On the other hand, stars form out of dense molecular clouds and after their formation may strongly radiate ultraviolet photons leading to photo-ionization and photo-dissociation of the surrounding molecular cloud. The interface between this ionized region around the star(s) and the dense molecular cloud is known as a Photo Dissociation Region \citep[PDR, e.g.][]{Hollenbach99} and have been extensively observed \citep[e.g.][]{Williams96,Bialy15,Larsson17,Tiwari18} as well as  theoretically modelled leading to insights on the various factors responsible for the atomic to molecular transition and vice versa  \citep[e.g.][]{Hollenbach99,Krumholz08,Wolfire10,Sternberg14,Bialy17}.

This association of atomic and molecular gas has mostly been explored using two easily observable tracers - the hyperfine transition of neutral atomic hydrogen, i.e the \HII \ line and various rotational transition lines from the \CO \ molecule. The atomic interstellar medium traced by the \HII \ line is thought to be composed of two components that are in approximate pressure equilibrium with each other - a Cold, $\sim$100 K, component with densities of 50 cm$^{-3}$ (the Cold Neutral Medium; CNM) and a warmer ($\sim8000$ K) component with much lower hydrogen number densities of 0.5  cm$^{-3}$ (the Warm Neutral Medium; WNM) \citep[e.g.][]{field65,cox05}.  The $\sim$100 K CNM around molecular clouds easily shows up in \HI~absorption against bright radio sources whereas \HI~emission is predominantly from the WNM [see \citet{HI_review} for a review of galactic HI studies].In spite of being an excellent tracer of the CNM, the \HII  \ line in absorption can get saturated at CNM conditions ($\tau_{\rm HI}=1$ at $\rm{N_H} =4.2 \times 10^{20}$cm$^{-2}$ for a typical CNM spin temperature of 100 K) making it difficult to study cold dense gas using the line. An alternate tracer of the Cold Neutral Medium are radio recombination lines arising from ionized carbon \citep{Salgado2017a}. Carbon with a first ionization energy of 11.2 eV is easily ionized in regions where Hydrogen is prominently neutral (ionization potential : 13.6 eV). The strong dependence of these lines on temperature, \emph{at least} $T^{-5/2}$, strongly biases the tracer towards the coldest regions. Also the low C-to-H abundance of $10^{-4}$ keeps these lines far from saturation. Apart from tracing the diffuse CNM \citep[e.g.][]{Palmer,Ananth}, { there is also evidence to suggest that Carbon Radio Recombination Lines (CRRLs) trace the molecular envelopes in { PDR \citep[e.g.][]{Sorochenko1996,salas2017b}}}. These properties make CRRLs excellent probes of the structure and physical condition of the CNM as well as allows us to understand the region of transition from neutral to molecular gas. 

A popular line of sight for ISM studies, especially CRRLs, have been that towards the bright radio source \casA \ {  (Cas~A), a 300 year old \citep[e.g][]{Thorstensen01} supernova remnant}. This line of sight cuts through the dense Perseus Arm of our galaxy and Cas~A provides a 6 arc-minute wide bright radio background to probe the intervening ISM against it. These factors have made this line of sight very popular with a variety of ISM studies being carried out [e.g. - \HII \ \citet{Bieging,Schwarz,Roy10}; \CO \ \citet{Liszt,Kilpatrick,Zhou18}; C{\sc I} \citep{Mookerjea06}; \HCO \ \citet{BatrlaFormal,reynoso}; \ch{NH3} \citet{BatrlaAmonnia}; \ch{OH} \citet{deJager}]. There has been significant debate in the literature on the location of Cas~A with respect to the intervening Perseus Arm gas and consequently its interaction with the gas \citep[e.g.][]{Liszt,Kilpatrick,Zhou18}. There is a broad consensus now that the gas is \emph{not} interacting with the supernovae remnant based on - (i) { Evidence that the distance between Cas A and the Perseus Arm gas is sufficiently large that it is not in contact with the expanding shell of the supernova} (see \citet{salas2017a} for an estimate of the distance between Cas~A and the Perseus Arm based on CRRL models and parallax measurements by \citet{Reed95,Choi14}), (ii) Lack of evidence for shock broadened gas \citep{Liszt,Zhou18}.

{ Recombination Line towards Cas~A was first detected by \citet{Konovalenko1980} where the line was misidentified as arising from Nitrogen and was later identified correctly as a CRRL by \citet{Blake1980}. Since then there have been multiple follow up studies of CRRLs towards the same line of sight \citep[e.g.][]{PayneCasA,Anantharamaiah1985,AnantharamaiahCasA,Kantharia98,Stepkin07}}. Recently, \citet{oonk,salas2017a,salas2017b} obtained new observations of CRRLs towards Cas~A using the LOFAR $\&$ the WSRT and performed a detailed modelling of these observations using the latest { departure coefficients} available from \citet{Salgado2017a,Salgado2017b}. \citet{oonk,salas2017a} presented a spatially unresolved study of the Perseus Arm gas towards Cas~A and found that the physical conditions of the gas resemble the CNM phase with the electron temperature T$_e = 69-98$ K. \citet{salas2017b} performed an arcmin scale resolved study of recombination lines towards Cas~A and compared them with molecular gas tracers towards Cas~A. These authors reported a non-uniform distribution of CRRL emission over the face of Cas~A. Specifically, they found evidence for an unresolved elongated structure over the southern region of Cas~A. Their parsec scale study also suggested that CRRLs trace the outer envelope of molecular gas in the Perseus Arm. 

In this paper, we present the first sub-parsec scale study of CRRL emission towards Cas~A using the upgraded Giant Metrewave Radio Telescope with the aim of understanding the spatial and { velocity structure} of the Perseus Arm gas traced by CRRLs and their connection to molecular gas. Section \ref{sec:obs} describes the uGMRT observations, section \ref{sec:dataanlysis} elaborates on the analysis techniques used to derive optical depth cubes from these observations, in section \ref{sec:results} we present the spatially integrated spectra along with the optical depth cubes, in section \ref{sec:discussion} we discuss various features of the CRRL emission and place them in the context of molecular gas towards the same line of sight and finally in section \ref{sec:conclusion} we summarize and conclude this work.
\section{Observations}
\label{sec:obs}
\begin{table*}[]
    \centering
    \begin{tabular}{c|c|c|c|c|c|c|c}
        \hline
        ID & Observation  & Frequency  & Channels & Bandpass  & Lines  & Channel & On-source \\
          & Date & Range &  &  Technique &  Cove &  Width & Time   \\ \hline
         I & June 2017 & 410-460 MHz & 16,384 & Position Switching & C$\alpha$244-C$\alpha$250 & 1.99 - 2.23  \kmps & 2 hours \\
         II & June 2017 & 410-460 MHz & 16,384 & Frequency Switching & C$\alpha$244-C$\alpha$250 & 1.99 - 2.23  \kmps & 5 hours\\
         III & March 2018 & 435.5-448 MHz & 16,384 & Frequency Switching & C$\alpha$245-C$\alpha$246 & 0.51 - 0.52  \kmps & 10 hours\\
    \end{tabular}
    \caption{Summary of observations using the upgraded Giant Metrewave Radio Telescope (uGMRT) of Carbon Radio Recombination Lines towards \casA.}
    \label{tab:obs}
\end{table*}
We made multiple observations of Recombination Lines towards \casA \  with the Giant Metrewave Radio Telescope \citep[GMRT,][]{Swarup91} (see table \ref{tab:obs} for a summary of observations). The first set of observations were carried out in June 2017 with a total telescope time of $\sim 12$ hrs. These were meant to test the capabilities of the upgraded GMRT \citep[uGMRT,][]{gupta17} in achieving high dynamic range spectra over a large bandwidth (50MHz) and hence involved trying out two different { bandpass calibration} strategies - (i) Position Switching (ii) Frequency Switching. We used the GMRT Wideband Backend (GWB) to observe Cas~A { in the 410-460 MHz frequency range} with 16k spectral channels (the maximum number of channels that the GWB could process at the time these observations were made) allowing us to cover 7 C$\alpha$n (n$\rightarrow$n-1) lines from C$\alpha244$ to C$\alpha250$ with a channel width of $\sim2$  \kmps, barely sufficient to resolve out the lines. For all our observations, 3C468.1 was used as the phase and amplitude calibrator for Cas~A. The position switching run was $\sim 6$ hours long and Cygnus A was used as the bandpass calibrator. Given that Cygnus A is almost as bright as Cas~A at our observing frequencies, we split the time equally between the source and the bandpass calibrator. We note that this led to at-least {  a factor of} $\sqrt{2}$ degradation in the signal-to-noise and also large observing time overheads with only $\sim 2$ hours on Cas~A. {For the other 6hr run, after every 5 minutes of observation on Cas~A, we switched the frequency of the local oscillator by $1$ MHz at the post-digitization stage.} In this particular run, the total on-source time was $\sim5$ hours.

A follow-up observation was made in March 2018 to obtain high velocity resolution spectra on two of the seven lines, C$\alpha245$ and C$\alpha246$, using the 12.5 MHz mode of the { GWB covering the 435.5-448 MHz frequency range} with 16k spectral channels. We switched the frequency by 1 MHz every 10 minutes (again in the post digitization stage) to bandpass calibrate the data. Cas~A was continuously tracked for $\sim10$ hours. Observing a phase calibrator was not required because a sufficiently high signal-to-noise ratio image of \casA \ made from the June 2017 observations was used to self-calibrate the data.  

\section{Data Analysis}
\label{sec:dataanlysis}
 For the purpose of this study, we extracted 256 channels around each C$\alpha$ line and all subsequent analysis were done on these pieces using the software package CASA \citep[version 5,][]{CasaRef}. In the first step, we flagged and calibrated each of these pieces for different observation runs separately. The preliminary calibration strategy followed was slightly different for each of the three observation runs and are described below.
 \subsection{Preliminary Calibration}
 \subsubsection{Low Resolution Data : Position Switching}
 The bandpass calibrator, Cygnus A (Cyg A), is resolved by the GMRT on most baselines and hence a model of Cyg A is required to find bandpass solutions.  The bandwidth of each of the 256 channel {sub-band} is sufficiently small ($\Delta\nu=781.25$ kHz; $\Delta\nu/\nu_0 \sim 2 \times 10^{-3} $) to neglect variation of { the source visibility on each baseline} as a function of frequency across the {sub-band}. With this assumption, we averaged 16 channels per baseline per time integration and copied the values over to the model column of the CASA measurement set. We then solved for the bandshape using the standard CASA task {\sc bandpass}. We note that this procedure takes care of both the source structure as well as the broadband variation of antenna gain with time and is similar to the "divide by channel zero" bandpass strategy available in classic AIPS. Phase and amplitude solutions were then derived from observations on 3C468.1 and applied on Cas~A along with the bandpass solutions. 
  \subsubsection{Low Resolution Data : Frequency Switching}
  \label{sssec:LRFS}
 As in the case of Cygnus A, \casA \ is also resolved on most GMRT baselines and we thus used a procedure similar to the one described in the preceding section to derive bandpass solutions from the frequency switched observations. But these bandpass solutions were neither effective in taking out the smooth variation of the band nor the rapid oscillations seen for few antennas. We thus decided not to bandpass calibrate the data at this stage since a polynomial fit to the continuum visibilites at a later stage of the analysis took care of the smooth variations. This was not the case for a few antennas which showed strong oscillations in the band and were flagged at this stage. Only phase and amplitude solutions derived from observations on 3C468.1 were applied on Cas~A. 
 
\subsubsection{High Resolution Data : Frequency Switching}
Bandpass solutions from the frequency switched data were generated in a way similar to that described in section \ref{sssec:LRFS}. The application of these solutions were necessary to take out some narrowband features in the { bandpass shape}. A single solution per antenna per spectral window was generated for the entire observation run and was found to be effective in taking out these features.
 
\subsection{Self-Calibration and Continuum Imaging}

\begin{figure}
    \centering
    \includegraphics[width=\columnwidth]{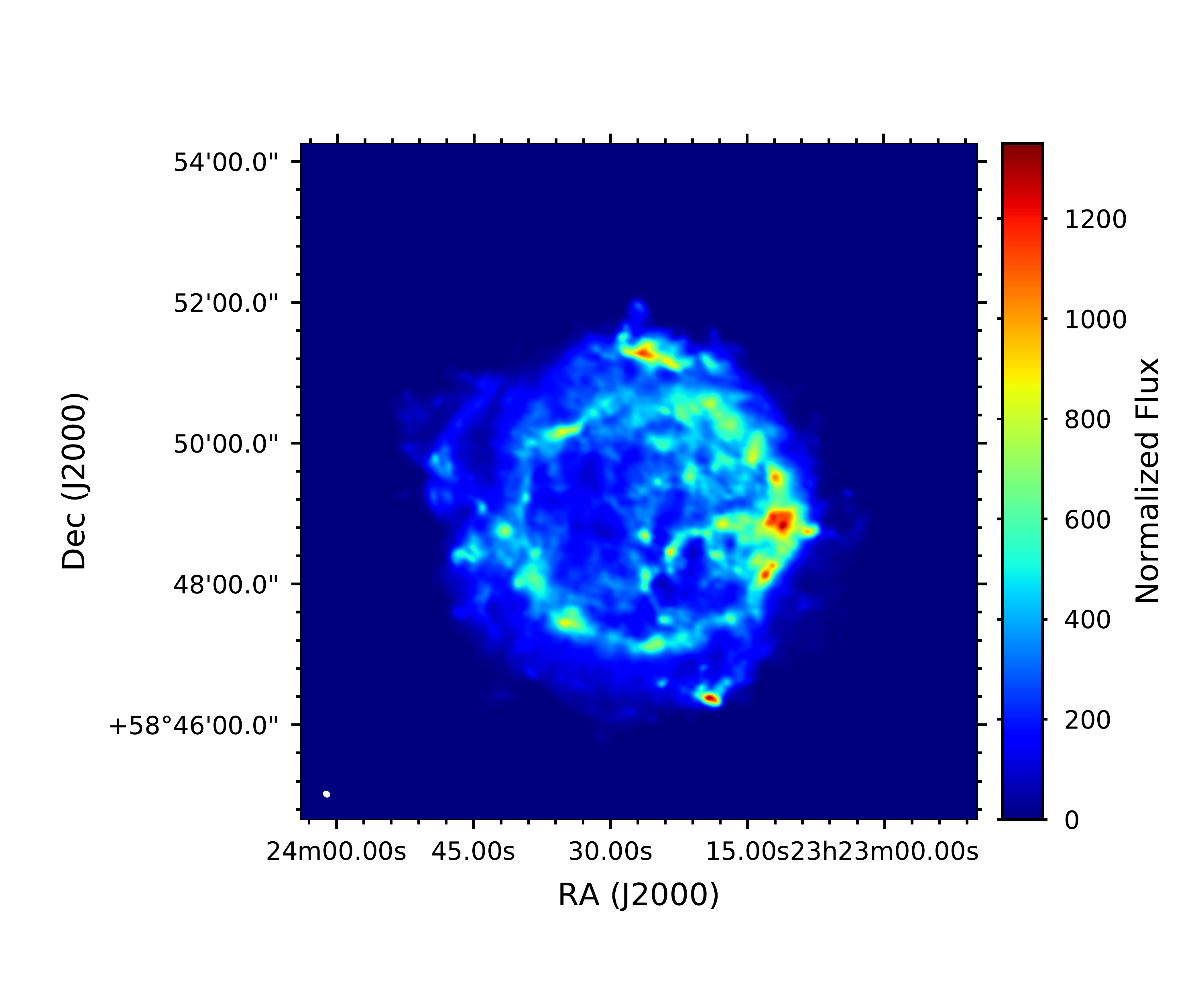}
    \caption{Continuum Image of \casA \ at a resolution of 5.3"$\times$4.3". The flux units have been normalized by the noise in the map. The circle in bottom left shows the synthesized beam.}
    \label{fig:CasA_cont}
\end{figure}
We did the self-calibration of the entire dataset (combining all three observations) in multiple stages. In the first stage, we combined both the coarse resolution calibrated dataset (from observations I \& II in table \ref{tab:obs}) around the C$\alpha246$ transition. We performed two rounds of phase only self-calibration followed by 3 rounds of phase and amplitude self-calibration on this combined data. A single gain solution per antenna was derived for each scan in the first selfcal iteration and then for every 120 s in all subsequent iterations. The imaging was done excluding the channels where the CRRL emission is expected (i.e. velocity space of both the Perseus Arm, -60 \kmps to -30 \kmps, and the local Orion, -7 \kmps to 7 \kmps, component) using the CASA task {\sc tclean} with uniform weighting,  w-projection and the multiscale clean algorithm; three scales were specified for the clean algorithm - zero, the size of the synthesized beam and five times the size of the synthesized beam. We note that none of our visibility data were flux calibrated implying that all continuum images derived were in arbitrary flux units. However, this arbitrary scale canceled out when we divided the line emission by the continuum emission to form the optical depth cubes.

In the next stage, we combined the preliminary calibrated dataset of  \emph{all lines from all observing runs} to form one visibility file. The model of Cas~A obtained in { the first stage, from observations around the C$\alpha246$ transition,}  was then used as an initial model for self-calibration of \emph{the whole dataset}. We did multiple rounds of imaging followed by phase and amplitude calibration till there was no improvement in the image dynamic range. {Imaging was done with line-free channels only and with similar parameters as in the first stage.} This procedure efficiently corrected for the spatially integrated spectral index of Cas~A by equalizing the total continuum flux in the frequency sub-bands around each C$\alpha$ line. At these observing frequencies, once the global spectral index for the continuum emission of Cas~A is corrected for, { the variation of the residual continuum flux at each pixel across the 50 MHz band is less than $2\%$ \citep{DeLaney}}. This step also normalizes the flux level of the various { C$\alpha$ lines} to the same continuum flux ensuring that the lines can be imaged together and then divided by the continuum image to obtain the optical depth cubes (see next section for more details on this aspect).  The final image obtained has a resolution of $5.3" \times 4.3"$ (see figure \ref{fig:CasA_cont}).

\subsection{Spectral Imaging}
The high resolution continuum image (shown in figure \ref{fig:CasA_cont}) was Fourier transformed to the visibility plane using the CASA task {\sc ft} and subsequently subtracted from the calibrated data using the CASA task {\sc uvsub}. This took care of all zeroth order effects in the spectral baseline. We fit third order polynomials to the residual of each baseline at every time integration using the task {\sc uvcontsub}; the polynomial fit was performed separately for each spectral window by excluding the velocity space of both the Perseus Arm, -60 \kmps to -30 \kmps, and the local Orion, -7 \kmps to 7 \kmps, component. We note that this continuum subtraction procedure was found to be adequate in taking out residual patterns in the bandpass of the data even for observations where no bandpass calibration was done.

\subsubsection{Stacking lines in visibility space}
Similar to previous studies of recombination lines towards Cas~A \citep[e.g.][]{oonk,salas2017b}, we wanted to stack all of the nearby lines together to boost { the signal-to-noise ratio}. The signal-to-noise ratio on the stacked lines for our observations are good enough that we can spatially resolve out these lines. The standard approach to this problem has been to image each line separately and then stack them in the image plane. This approach has the disadvantage that the clean algorithm will be unable to pick up low level flux from each cube. {Both these low level fluxes  and  the residual deconvolution error will add up and become statistically significant once the lines have been stacked}. To avoid this issue \emph{we stack the lines in the UV plane}. We do this in CASA in the following way - (i) The frequency of each channel was scaled to align all the lines at the rest frame frequency of C$\alpha$247, (ii) The baseline coordinates (in m) were scaled by the inverse of this factor to ensure that CASA computes the UVW points properly during imaging. 
 
 This \emph{uv-stacked} visibility data was used to produce spectral cubes. The spectral imaging was done using the CASA task {\sc tclean} using the Högbom clean algorithm (which identifies clean components during each minor cycle)  with Briggs weighting of 0.5 and appropriate uvtapers to control the spatial resolution. Table \ref{tab:cubes} lists the three spectral cubes that were produced with different spatial and spectral resolution. Optical depth cubes were subsequently produced by dividing these line cubes by the continuum image smoothed to the same resolution. Pixels with low continuum flux were masked to avoid high optical depth rms at the edge of the continuum emission. 
\begin{table}

 \begin{tabular}{c|c|c|c}
 \hline
 Cube & Lines & Beam  & Spectral Resolution  \\ \hline
 I & C$\alpha$244-C$\alpha$250 & 18", 0.29pc & 4.40  \kmps \\
 II & C$\alpha$245-C$\alpha$246 & 20", 0.32pc & 0.55  \kmps \\
 III & C$\alpha$245-C$\alpha$246 & 30", 0.48pc & 0.55  \kmps \\
 \end{tabular}
 \caption{Optical depth cubes used in this work. The beam-width is converted to linear distance using an assumed distance of 3.3 kpc to the Perseus Arm cloud \citep{salas2017a}}
 \label{tab:cubes}
 \end{table}
\section{Results}
\label{sec:results}
\subsection{Spectra}
\begin{figure}
    \centering
    \includegraphics[width=\columnwidth]{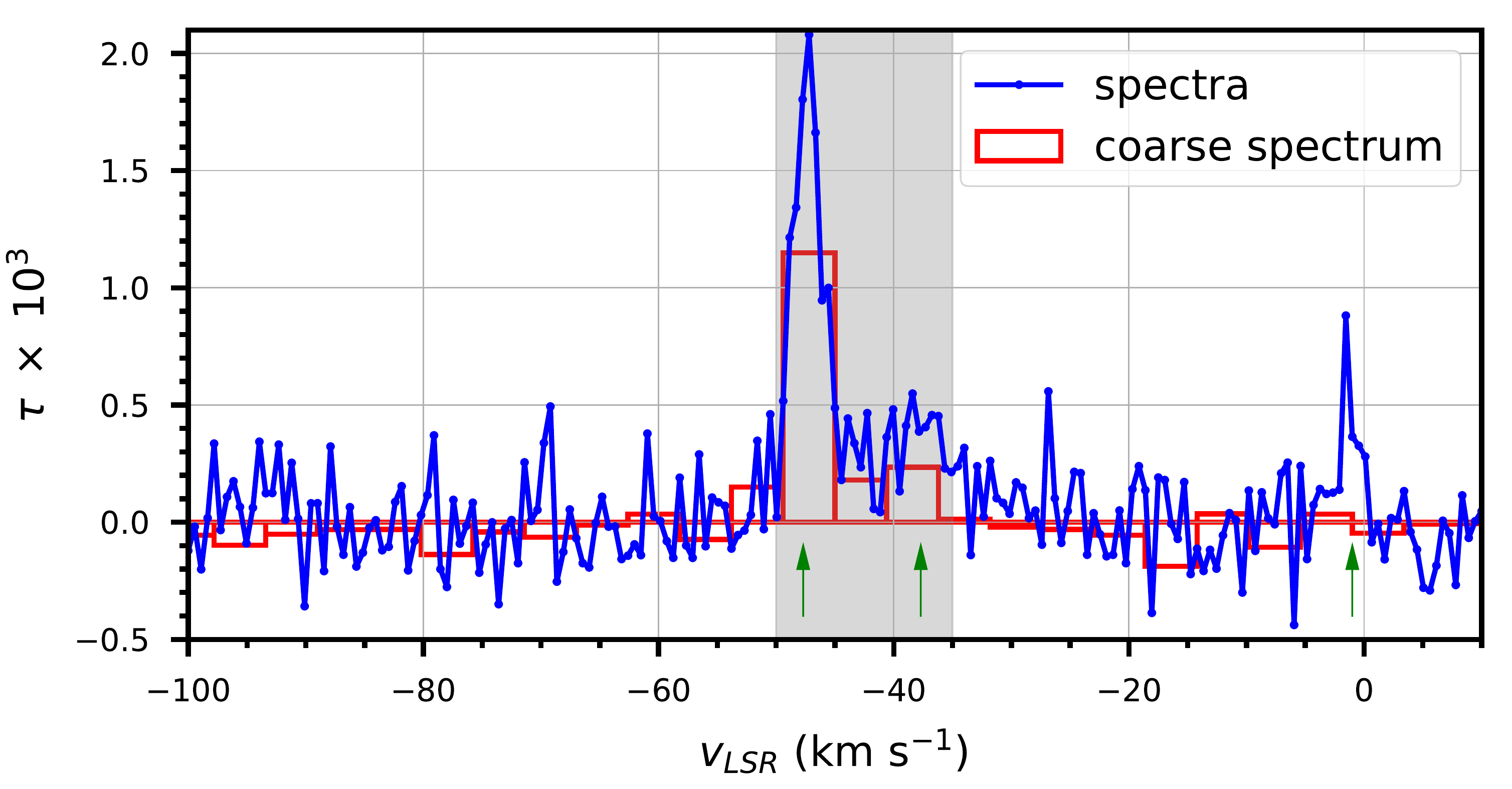}
    \caption{Spatially integrated optical depth spectra towards Cas~A. The blue curve shows the 0.55  \kmps stacked spectra from the C$\alpha$245-C$\alpha$246 observation whereas the red bar plot shows the 4.4  \kmps stacked spectra obtained from observations of C$\alpha$244-C$\alpha$250. The shaded region marks the Perseus Arm velocity range. The arrows mark the three components detected in most atomic and molecular line studies towards Cas~A at -47  \kmps, -35 to -42  \kmps and 0  \kmps.}
    \label{fig:spectra}
\end{figure}
The line of sight towards Cas~A shows multiple velocity components in most atomic and molecular gas tracers. The are three main features - (i) A strong feature peaked at -47  \kmps, (ii) Multiple blended components between $\sim -35$ \kmps and $-42$ \kmps and   (iii) A feature near 0  \kmps. The first two features are associated with gas in the Perseus Arm and { the 0 \kmps component with gas in the local Orion Spur component} \citep[e.g.][]{Kantharia98,Liszt,reynoso,Kilpatrick,oonk}.

We computed the global spectrum of CRRLs from our observations by spatially averaging over the optical depth cubes. The high velocity resolution spectra shown in figure \ref{fig:spectra} is from the C$\alpha$245-C$\alpha$246 observations (cube III in table \ref{tab:cubes}) and clearly shows the Perseus Arm components between -50  \kmps and -35  \kmps as well as the local Orion spur component near 0  \kmps. Figure \ref{fig:spectra} also shows the coarse resolution spectra of C$\alpha$244-C$\alpha$250 (extracted from cube I in table \ref{tab:cubes}) where both the Perseus Arm components are detected but we note that the lines are barely resolved. 

\subsection{The Spatial and Velocity Structure of CRRL emission}
\begin{figure}
    \centering
    \includegraphics[width=\columnwidth]{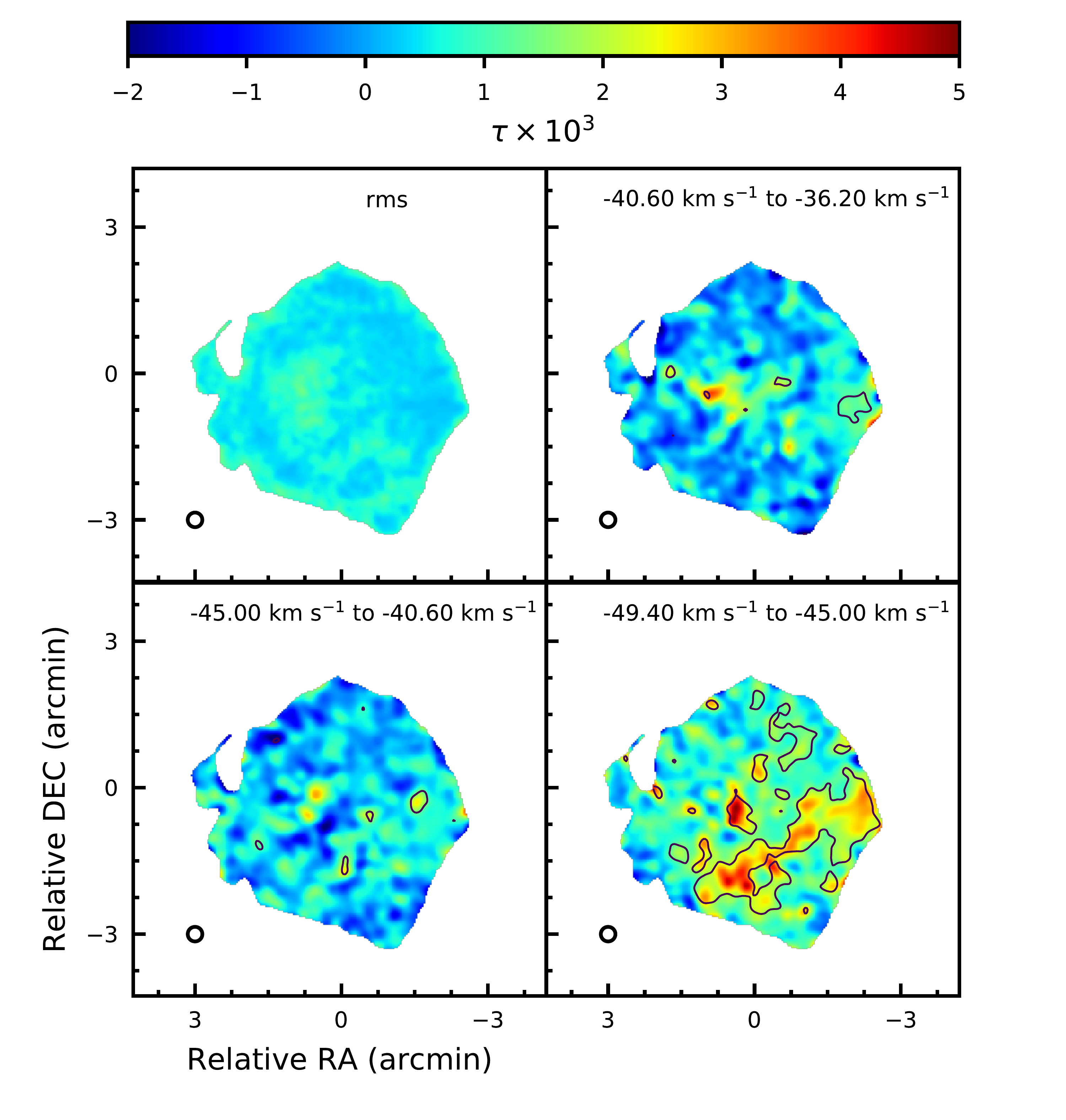}
    \caption{Velocity integrated optical depth maps of recombination line emission from the Perseus Arm, C$\alpha$244 - C$\alpha$250, at a spatial resolution of 18 arc-seconds (linear scale of 0.29 pc). The top left panel shows the variation of optical depth rms over the face of Cas~A. The black contours mark the region where the lines are detected with a signal-to-noise greater than 4.}
    \label{fig:lowresCube}
\end{figure}
\begin{figure*}
    \centering
    \includegraphics[width=2\columnwidth]{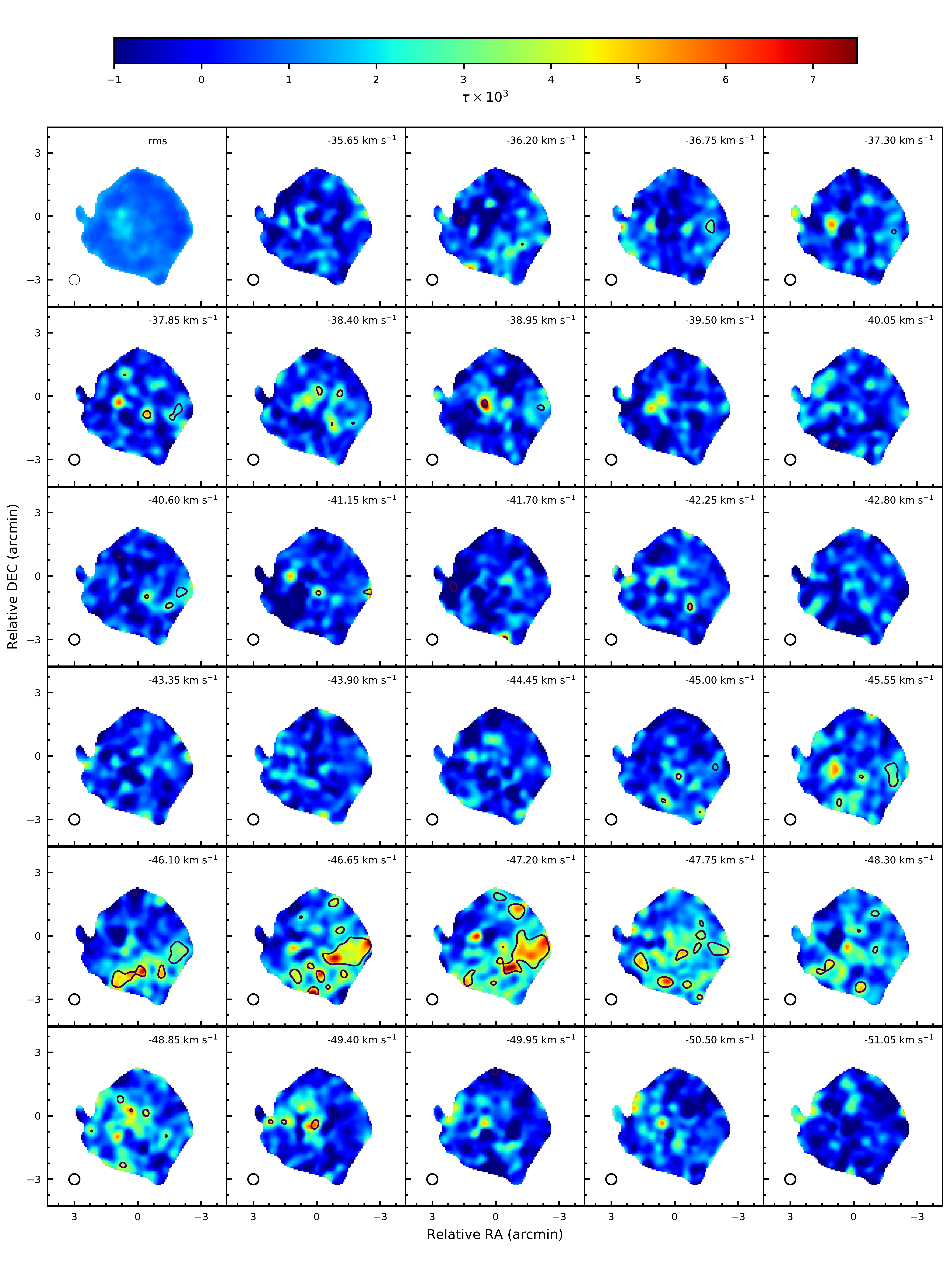}
    \caption{Optical depth maps of recombination line emission from the Perseus Arm, C$\alpha$244 - C$\alpha$250, at a spatial resolution of 30 arc-seconds (linear scale of 0.48pc) and a velocity width of 0.55  \kmps. The top left panel shows the variation of optical depth rms over the face of Cas~A. The black contours mark the region where the lines are detected with a signal-to-noise greater than 4.  }
    \label{fig:highresCube}
\end{figure*}

Slices of the stacked optical depth cube of C$\alpha$244-C$\alpha$250  (cube I in table \ref{tab:cubes}) in the Perseus Arm velocity range (-35 \kmps to -51 \kmps) are shown in figure \ref{fig:lowresCube}. { These channel maps are at a spatial resolution of 18" corresponding to a linear scale of 0.29 pc with an assumed distance of 3.3 kpc to the gas clouds \citep{salas2017a}. The maps provide, for the first time, a sub parsec view of the structure of CRRL emission from the Perseus Arm.} The component centered at -47  \kmps shows an elongated, \emph{bridge-like}, structure from { south-east} to { north-west}. This feature was also detected by \citet{salas2017b} in their 70" maps of CRRLs towards Cas~A but their coarse spatial resolution did not allow them to resolve the sub-structure that we see here. Our high resolution map shows significant sub-pc structure along this \emph{bridge}; for example, note the unresolved knots of emission towards the southern direction. We also detect some emission north of this \emph{bridge} in the component centered at -47  \kmps. The remaining velocity components of the Perseus Arm show clumpy emission in our maps. The emission centered at -38  \kmps has a diffuse component to the { west} and an emission knot towards the center of Cas~A. These features are similar to the features seen in molecular gas towards Cas~A \citep[e.g.][]{Liszt} and will be discussed further in section \ref{ssec:mol_gas}. {The rich spatial sub-structures detected in our CRRL maps are not seen in the other tracer of the CNM, the \HII \ line in absorption, because it is saturated in most regions where we see a high CRRL optical depth.}

We use our C$\alpha$245-C$\alpha$246 observations (cube I in table \ref{tab:cubes}),  which have a better velocity resolution, to understand the velocity structure of these emission components. The stacked channel maps with a channel width of 0.55  \kmps are shown in figure \ref{fig:highresCube}. These maps have a larger beam-width of 30" (linear scale of 0.48 pc) which gives a better signal-to-noise than the maps at higher resolution and are hence better suited to study the velocity profiles. These maps provide us with insights on various components of the CRRL emission from the Perseus Arm : 
\begin{itemize}
    \item The component centered at -47  \kmps shows a drift from { south-west to north-east} with increasing negative velocity. The maps at -49.40  \kmps and -48.85  \kmps show emission primarily from the { north-east} region whereas the bridge like structure along the { south-west} is seen between -48.30  \kmps and -45.55  \kmps. This velocity gradient is also seen in 21cm absorption maps towards Cas~A \citep{Bieging} which is indicative of CRRL emission tracing the diffuse Cold Neutral component of the ISM. \citet{Bieging} noted that the velocity gradient is along the direction of constant galactic latitude and might be streaming motion along the spiral arm. 
    
    \item We see another velocity gradient along the elongated \emph{bridge} that is detected between -46.10  \kmps and -47.20  \kmps with the { eastern} component peaking at -46.10  \kmps and the { western} component peaking at a more negative -47.20  \kmps. 
    \item The emission between -36.75  \kmps and -45.00  \kmps shows multiple components localized primary towards the center of Cas~A (For example, see the channel map at -38.95  \kmps) and also towards the { western} hotspot (e.g. the component detected at -40.60  \kmps).  This further establishes that the emission from this region of the velocity space is a blend of multiple, possibly independent, components. 
    \item The velocity maps show features that appear on a single channel and are not detected in adjacent channels. We explore the implication of such components in section \ref{ssec:single_chan}.
    
\end{itemize}

In summary, we find evidence for rich structures both in the spatial directions as well as the velocity direction establishing that the CRRL emission shows significant deviation from the simple uniform slab model used to derive physical properties of the Perseus Arm gas in multiple studies \citep[e.g.][]{oonk,salas2017b}. 

\section{Discussion}
\label{sec:discussion}
\subsection{Evidence for multi-component gas}
\label{ssec:multi-comp_gas}
\begin{figure*}
    \centering
    \includegraphics[width=2\columnwidth]{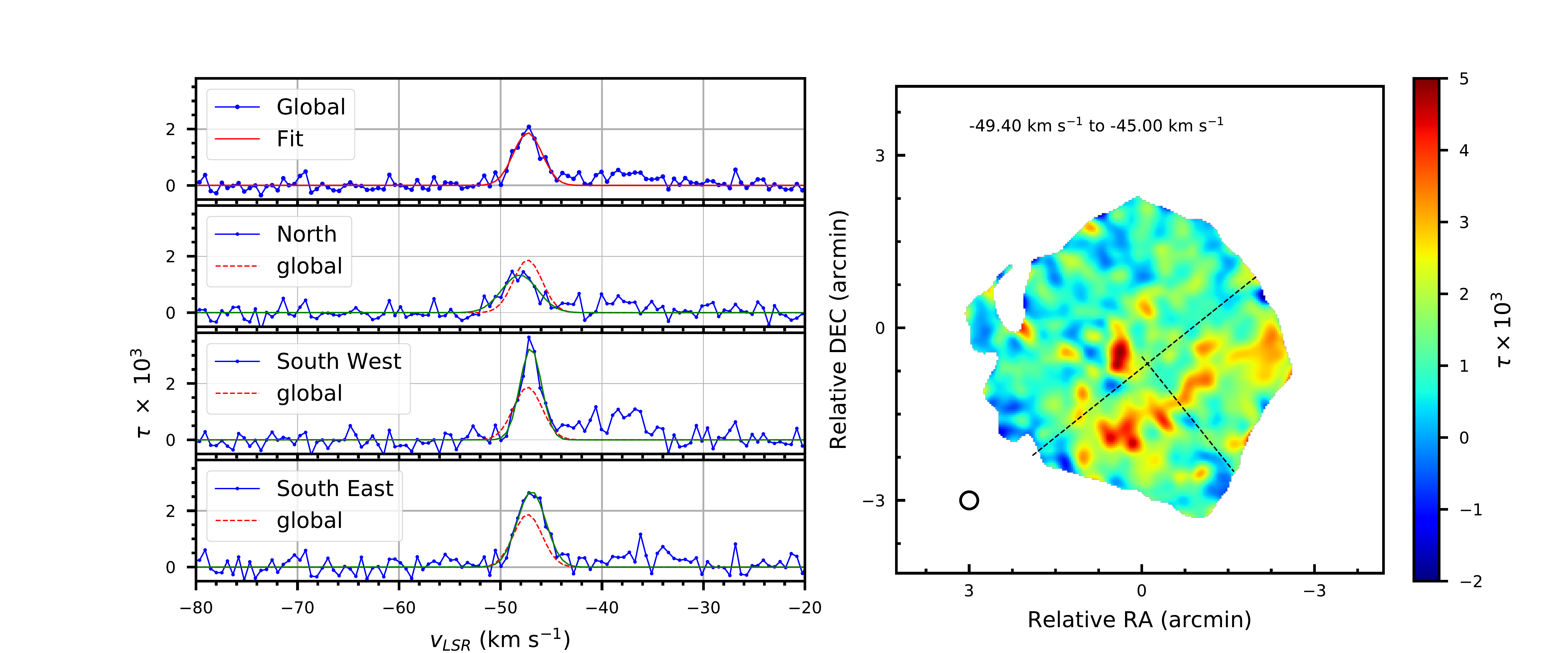}
    \caption{Spatially integrated optical depth spectra extracted from various regions over the face of Cas~A. The right panel shows the -47  \kmps channel map at a  spatial resolution of 18" (linear scale of 0.29 pc) overlaid on which are the boundaries separating these regions. The left panels show the spectra extracted over - (i) Entire face of Cas~A, (ii) North of the  \emph{bridge}, (iii) { Western} part of the \emph{bridge} { (South West)}, and (iv) { Eastern} part of the \emph{bridge} { (South East)} [top to bottom]. The red curve in each panel shows the line profile fit to the global spectra for comparison.}
    \label{fig:multipleRegions}
\end{figure*}
\begin{table}

 \begin{tabular}{c|c|c|c|c}
 Region & $v_0$  & FWHM  & $\tau_0 $ & $\int{\tau d\nu}$  \\
\; & ( \kmps)& ( \kmps)&($10^{-3}$)&(Hz)\\\hline 
 Global &  $-47.32 \pm 0.11$ & $3.31 \pm 0.26$ & $2.9 \pm 0.2$ & $9.6 \pm 1.4$\\ 
 North &  $-48.10 \pm 0.19$ & $4.16 \pm 0.45$ & $2.1 \pm 0.2$ & $8.9 \pm 1.7$\\
 { South West} &  $-47.02 \pm 0.09$ & $2.47 \pm 0.21$ & $5.2 \pm 0.4$ & $12.6 \pm 2.0$\\
 { South East} &  $-46.94 \pm 0.26$ & $3.28 \pm 0.26$ & $4.2 \pm 0.3$ & $13.7 \pm 2.0$\\

 \end{tabular}
 \caption{Parameters of Gaussian profile fits to the spectra shown in figure \ref{fig:spectra} extracted from various regions over the face of Cas~A. The columns list the line center ($v_0$), the full width at half maxima of the line profile (FWHM), the peak optical depth ($\tau_0$) and the integrated optical depth ($\int{\tau d\nu}$). }
 \label{tab:line_params}
 \end{table}
The Perseus Arm component centered at -47  \kmps has been treated in the literature as a single component gas cloud \citep[e.g.][]{KanthariaCasA,salas2017a} whereas the maps presented in figure \ref{fig:highresCube} show evidence for significant spatial and spectral sub-structure. To investigate further the nature of these components we split the emission over the face of Cas~A into three parts along the lines shown in the right panel of figure \ref{fig:multipleRegions}  covering (i) The { western} part of the \emph{bridge}, (ii) The { eastern} part of the \emph{bridge}, (iii) Emission north of the \emph{bridge}. The high velocity resolution spectra extracted from each of these regions along with the spectra over the entire face of Cas~A are shown in figure \ref{fig:multipleRegions}. The pressure and radiation broadening of CRRLs are negligible for these C$\alpha$ lines \citep{oonk} and hence we approximate the Voigt line profile as a Gaussian and fit to the spectra extracted from each region. The line profile parameters are listed in table \ref{tab:line_params}. 

The spectra integrated over the entire face of Cas~A has an FWHM of $3.31 \pm 0.26$  \kmps which is consistent with the findings of \citet{oonk}. The emission from the region north of the \emph{bridge} is very different from both the components on the bridge - it shows a lower optical depth and a broader line profile which is centered at a more negative velocity. The broader line profile may be indicative { of a warmer, more diffuse component} which is consistent with the fact that there is no detection of any molecular tracers from the northern region in the -47  \kmps component of the Perseus Arm (see section \ref{ssec:mol_gas} for comparison with molecular tracers). {Also, \citet{salas2017b} found a higher electron temperature of the gas over this northern part of \casA.}  { The properties of the emission from the \emph{bridge}} are also very different in the south-east and south-west regions. The { south-west} emission shows a narrower and stronger line profile compared to the { south-east} emission. These differences provide further evidence that the -47  \kmps carbon recombination line emission is not homogeneous and is in-fact made up of multiple components with different emission characteristics.

\subsubsection{Narrow Emission Lines}
\label{ssec:single_chan}
\begin{figure}
    \centering
    \includegraphics[width=\columnwidth]{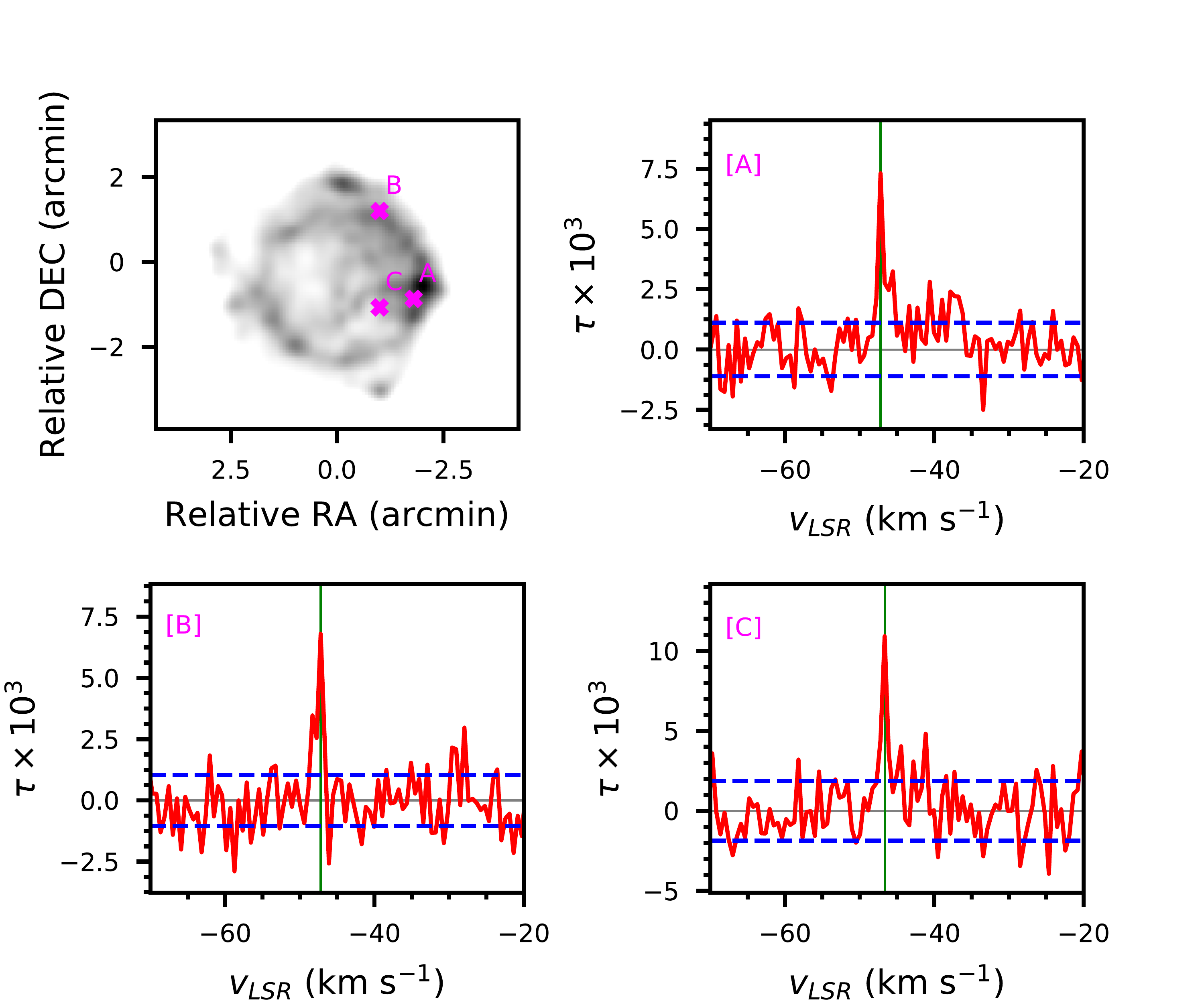}
    \caption{Detection of features with narrow line width. The top left panel shows the location of these features overlaid on the continuum map of Cas~A at a resolution of 20". The remaining three panels show the spectra of these features in red. The blue dotted line marks the 1$\sigma$ noise level in these spectra and the green vertical line indicates the velocity at which the lines are detected (-47.2  \kmps for all three spectra). }
    \label{fig:single_chan}
\end{figure}

Motivated by the apparent spatial shift of features in neighbouring channels separated by 0.55  \kmps (see figure \ref{fig:highresCube}), we searched the high resolution optical depth cube (cube II in table \ref{tab:cubes}) for features that appear only in a single channel. We performed the search by looking for features that are significant in one channel at n$\sigma$ but are less than n$\sigma$/2 significant in the adjacent channels. The threshold, n was determined by varying it till all detection fall in the velocity range of the Perseus Arm; this minimizes the probability of incorporating a deviant noise feature in our sample. We detected three such single channel emission spots, all centered at -47.2  \kmps, with a signal-to-noise greater than 6 {(i.e. n > 6)} in at least two adjacent spatial pixels. The spectra along with their locations over the face of Cas~A are shown in figure \ref{fig:single_chan}. 

We derived a $99\%$ confidence upper-limit on the full width at half maxima of these lines by assuming a Gaussian line profile. These upper limits for the FWHM are 1.4  \kmps, 1.5  \kmps and 1.75 \kmps for the three detected features A, B and C respectively. These velocity widths are large compared to the thermal broadening of 0.62  \kmps at typical CNM temperature of 100 K. But these are significantly smaller compared to the FWHM of the spatially integrated line of 3.31 $\pm$ 0.26  \kmps and would be further indication of the gas cloud at -47  \kmps being made up of multiple components. 

Molecular clouds have been long known to follow a size-velocity width relation (Larson's Law) which is thought to be a manifestation of turbulence driven cloud fragmentation \citep{Larson81}. There is now evidence to support that there can be a break in this relationship below 0.1 pc where one observes "coherent clouds" with constant near thermal velocity dispersion inside them \citep{Pineda10,Chen18}. The scale at which we observe the narrow line features (0.32 pc) is close to this break scale and it is possible that the CNM also follows { a turbulence driven size-velocity dispersion relationship where the velocity dispersion decreases with scale, until one reaches the Kolmogorov dissipation scale. At this scale, the turbulent energy cascade becomes sub-dominant, and the velocity width is equal to the thermal velocity of the gas} \citep[for example][provides evidence for such a power law scaling using \HII \ studies]{Roy08}. Unfortunately the signal-to-noise in our current data { does not allow us to investigate} this hypothesis further and can be part of a future work. 
\subsection{Relation to molecular gas}
\label{ssec:mol_gas}
Molecular gas towards \casA \ has been studied in great detail using a wide variety of tracers - $^{12}$\CO $\&$ $^{13}$\CO \  \citep{Liszt,Kilpatrick,Zhou18}, \HCO \ \citep{BatrlaFormal,reynoso}, \ch{NH3} \citep[e.g.][]{BatrlaAmonnia} and \ch{OH} \citep[e.g.][]{deJager}. {There is also some evidence to suggest that recombination lines trace the outer envelopes of dense molecular gas viz. a stratification wherein ionized carbon (\ch{C+}) gets converted to neutral carbon and molecules in dense cores of molecular clouds \citep[e.g.][]{Sorochenko1996,salas2017b}}. Hence one would expect a correspondence between the distribution of molecular tracers and that of carbon recombination lines. In this study, we compare our CRRL maps with the diffuse molecular gas traced by \CO \ and the clumpy dense molecular gas traced by Formaldehyde (\HCO). 

\begin{figure*}
    \centering
    \includegraphics[width=2\columnwidth]{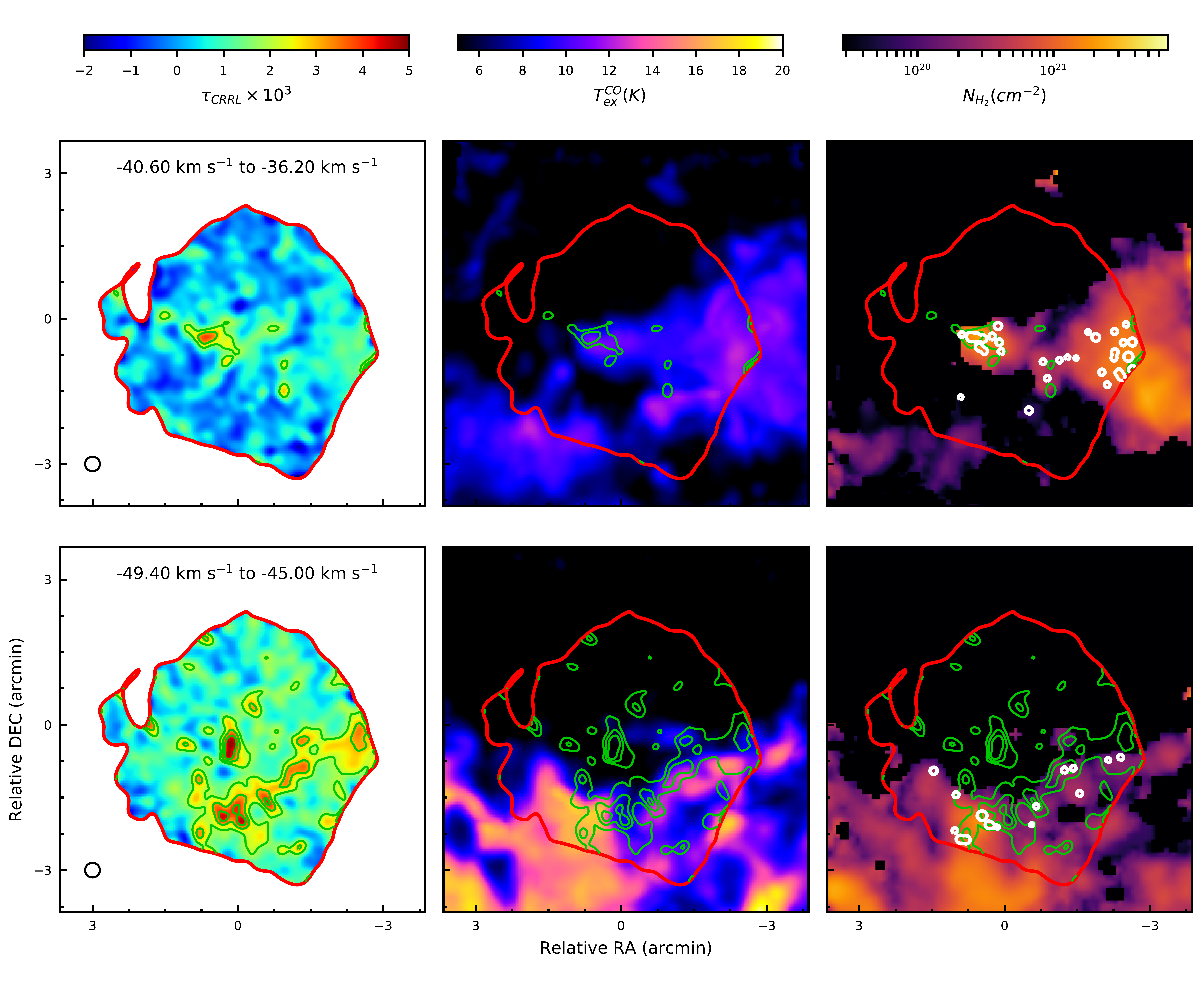}
    \caption{Relation of CRRL emission to molecular gas tracers. The panels show from left to right (i) CRRL emission, (ii) Excitation temperature of \CO \ ($\rm{T_{ex}}^{\CO}$), (iii) \Htwo \ column density derived from \CO \ maps ($\NH$). The \CO \ maps used to derive the physical parameters are from \citet{Zhou18}. The two different rows map the tracers in the two Perseus Arm component in the velocity range (i) -40.0 to -36.2  \kmps (top) (ii) { -49.4 to -45.0}  \kmps (bottom). All maps are at a resolution of 18" corresponding to a linear scale of 0.29 pc (beam marked in black circle). The red outline marks the edge of the continuum emission from Cas~A over which CRRLs are mapped. The green contours are of CRRL optical depth with the levels set at $\tau=2,3,4 \times 10^3$. Finally, the white circles mark the position of \HCO \ ($1_{11}$-$1_{10}$) clumps from \citet{reynoso}.}
    \label{fig:molecular_gas}
\end{figure*}

We used high resolution ($\sim$11") $^{12}$\CO (J=2-1) and $^{13}$\CO (J=2-1) maps from \citet{Zhou18} and convolved the CO maps to a spatial resolution of 18" and velocity resolution of $4.4 $ \kmps (the resolution of cube III). We then used these maps to derive the excitation temperature of \CO \ ($\rm T_{ex}^{CO}$) and the column density of \Htwo \ ($\NH$) by assuming a constant excitation temperature across the two species and a $^{13}$\CO \ to H$_2$ abundance of [$^{13}$CO]/[H$_2$]=$2\times10^{-6}$ \citep{Dickman78}\ (see appendix \ref{sec:appnxA} for details of the procedure). We used the list of \HCO \ ($1_{11}$-$1_{10}$) clumps detected by \citet{reynoso} to locate high density molecular gas over the face of Cas~A. Figure \ref{fig:molecular_gas} shows the distribution of C$^{+}$ as traced by CRRLs, diffuse molecular gas traced by \CO \ and dense molecular clumps traced by \HCO.

The \emph{bridge} between -49.40  \kmps and -45.00  \kmps is clearly visible in the CO maps. This \emph{bridge} is an extension of a large ($\sim$ 20') molecular complex that lies to the south of Cas~A \citep{Zhou18}. { The \Htwo \ column density derived from \CO \ observations along the bridge is in the range $\NH\approx1-4 \times 10^{20}$ cm$^{-2}$ except towards the { south-eastern} region which extends into a high column density molecular gas cloud beyond the southern edge of Cas~A with $\NH$ as high as $\NH\approx2 \times 10^{21}$ cm$^{-2}$. On the other hand, the excitation temperature over the \emph{bridge} is between $8-18$ K and is comparable to that found in the dense southern extension of the cloud.} This elongated feature seen in the recombination line maps overlaps well with the CO feature but there is possibly an offset wherein the recombination line peaks slightly north of the CO \emph{bridge}. All, except one,  \HCO \ clumps identified by \citet{reynoso} lie slightly to south of the CRRL \emph{bridge}. {These authors found high \Htwo \ column densities along these clumps of up to  { $\NH \sim 10^{21}$ cm$^{-2}$} which is significantly smaller than the typical $\NH \sim10^{20}$ cm$^{-2}$ diffuse structure traced by CO.} It is interesting to note that there is barely any molecular gas tracers detected to the north of the \emph{bridge}; this is also true for other molecular tracers not shown in this paper (for example, \citet{BatrlaAmonnia} detected Ammonia only in a subset of their pointings and none of them are to the north of the \emph{bridge} for this Perseus Arm component). { This is consistent with the picture that the gas in the northern part is likely to be warmer and more diffuse than the gas on the \emph{bridge}} (see section \ref{ssec:multi-comp_gas} where this component is shown to have wider line profiles). The distribution of molecular gas viz. CRRL emission is indicative of { a PDR} like chemical stratification from north to south wherein there is a gradual transition from ionized carbon to dense molecular complexes. { This picture is different from the PDR distribution proposed by \citet{salas2017b} where they suggest that ionized photons are incoming from the { north-west} towards the \emph{bridge}}. The existence of dense \HCO \ in the { western part} of the \emph{bridge} would be inconsistent with such a picture. 

{The spatial offset between the CRRL and the CO bridge is marginally resolved in our 18" (0.29 pc) resolution images. If we assume an edge-on PDF model with photon influx perpendicular to the bridge, and follow Salas et al. (2018) in assuming that CO is shielded from photo-ionization at $A_{FUV} \sim 1$, i.e. corresponding to $N_{H} \sim 1.15 \times 10^{21}$~cm$^{-2}$, we can use the measured offset between the CRRL and CO emitting regions to make an approximate estimate on the hydrogen density. The value we get, $n_H \sim  10^3$~cm$^{-3}$, is within a factor of a few
of the numbers derived from detailed modelling by \citep{oonk,salas2017b}.}

The diffuse molecular gas in the other Perseus Arm component between -40.60  \kmps and -36.20  \kmps is distinctly detected in two different regions - one towards the center of Cas~A and the other to its { west}.  The typical \Htwo \  column density traced by CO is significantly higher than in the -47 \kmps gas with $\NH \gtrsim 10^{21}$ cm$^{-2}$ with a $\rm{T_{ex}}^{\CO}\approx8-11$ K. There are a large number of formaldehyde clumps that are detected within the diffuse molecular gas. The formaldehyde clumps towards the center of Cas~A have large \Htwo \ column densities of $\NH\approx10^{22}$ cm$^{-2}$ associated with them \citep{reynoso}. We detect CRRL emission from both the regions - the central region is prominent in figure \ref{fig:multipleRegions} whereas the emission from the { western} region is more diffuse and is detected over a larger spatial scale. It is curious that the central emission region with extremely high molecular column densities is detected in CRRL emission - a likely explanation would be a projection effect wherein what we are seeing is the line of sight integrated \ch{C+} envelope of the molecular cloud. 

In summary, CRRL emission is always detected from molecular features in the Perseus Arm gas and this may be associated with the envelopes of these molecular clouds. But the converse is not true wherein we detect CRRL emission from the North of the \emph{bridge} with no associated detection in molecular tracers. { These may be more diffuse CRRLs associated with the Cold Neutral Medium which are also seen in \HII \ absorption maps towards Cas~A \citep[e.g.][]{Bieging}}
\section{Conclusion}
\label{sec:conclusion}
In this paper, we have presented the first sub parsec study of Carbon Radio Recombination Lines. These lines (C$\alpha$244-C$\alpha$250), observed using the GMRT, allowed us to probe the cold gas in the Perseus Arm against the bright radio source \casA. { High spatial resolution maps of these lines} were made from GMRT observations using a new technique of stacking the observed lines in the visibility space. This technique enabled effective deconvolution and imaging of observed lines using the CLEAN algorithm and allowed us to probe the spatial structure of CRRL emission  over a $\sim6$ pc region down to linear scales of 0.3 pc.  Our main conclusions are : 
\begin{enumerate}
    \item CRRL emission detected from the Perseus Arm in the velocity range of $\sim -50$  \kmps to $\sim -30$  \kmps shows diverse spatial structures down to our resolution scales (0.29 pc). The emission from the component peaked at $-47$  \kmps is  concentrated along a narrow ($\sim 3$pc) elongated structure (\emph{bridge}) along the south of Cas~A which shows further substructure that are unresolved in our observations. {We also detect emission north of this structure that peaks at a more negative velocity. Such a south to north velocity gradient is clearly seen in \HII~maps towards Cas~A.}  The other components centered at $\sim -37$  \kmps also show rich spatial structures. 
    \item We find evidence for changing velocity structures in the -~47  \kmps emission component across the face of \casA. Specifically, we find evidence for different velocity widths for the { western and eastern} part of the bridge ($2.47\pm0.21$ \kmps and $3.28\pm0.26$ \kmps respectively) and the emission originating from the north of the \emph{bridge} to have a broader velocity FWHM of $4.16 \pm 0.45$  \kmps . The components also show differences in peak optical depth of up to a factor of $1.8$. These varied velocity as well as spatial structures are indicative of the -47  \kmps gas to be a blend of multiple components with possibly different physical properties which got averaged out in past studies which treated the gas as a uniform slab over the face of Cas~A. 
    
    \item We detect emission regions in our 20" (physical size of 0.32 pc) optical depth cube that are detected only in a single 0.55 \kmps  velocity channel. We find a $99\%$ confidence upper limits of 1.4  \kmps, 1.5  \kmps and 1.75  \kmps on the width (FWHM) of these lines. These FWHM values are significantly narrower than the width of the spatially integrated line ($3.31\pm0.26$  \kmps) and are further indicative of a multi-component gas.
    
    \item We present a comparison of the spatial distribution of CRRL emission and that of molecular gas tracers towards the same line of sight. We investigate the relation between CRRL emission and the diffuse molecular gas traced by \CO \ rotational transitions as well as the high density clumpy molecular gas traced by the 6 cm \HCO \ line. The main results from this study are :
    \begin{itemize}
        \item  We find that an elongated \emph{bridge} is also detected in CO at $-47$  \kmps and is well aligned to the structure seen in CRRL maps. We find no detection of CO lines from the North of this bridge where the CRRL profile is broader.
        \item Most dense molecular clumps traced by the H$_2$CO line is seen to the south of the bridge. We propose that the $-47$  \kmps gas is a PDR with influx of photons from the northern direction. 
        \item In the other Perseus Arm component centered at -37  \kmps, we find that regions which are detected in CRRL emission are also detected in CO with very high column density ($\NH \approx 10^{22}$ cm$^{-2}$) formaldehyde clumps embedded inside. This suggests that these CRRLs trace the outer envelope of dense molecular clumps.
    \end{itemize}

\end{enumerate}
\section*{Acknowledgements}
We thank the staff of the GMRT who have made these observations
possible. GMRT is run by the National Centre for Radio Astrophysics
of the Tata Institute of Fundamental Research. We would also like to thank Nissim Kanekar for many useful discussions on the work presented in this paper.




\bibliographystyle{mnras}

\bibliography{bibliography}



\appendix

\section{Deriving Optical Depth and Excitation Temperature from CO Line Intensities}
\label{sec:appnxA}
\label{appndx:CO}
In the following calculation, we assume $\rm T_{ex}$ to be constant between both the species as well as between various J levels. This is strictly true only if the levels are thermalized, in which case everything is coupled to the kinetic temperature $T_{K}$. We also assume that the $^{12}\CO$  (J=2-1) line is optically thick such that $1-e^{-\tau} \approx 1$.  Verifying these assumptions is non-trivial given the uncertainty in how efficient is radiative trapping in these clouds.

Assuming a value for the $^{13} \CO$ to \Htwo \ abundance (X($^{13} \CO$)),  the \Htwo \ column density ($\NH$) is related to the integrated optical depth of the $^{13} \CO$ (J=2 to 1) transition ($\tau^{\rm{int}}_{13}$) and the excitation temperature ($\rm{T_{ex}}$) as 
\begin{equation}
\label{eq:13_upper}
\begin{split}
     \NH = & 1.65 \times 10^{20} \rm{cm}^{-2} \ \tau^{\rm{int}}_{13} \ \ \frac{ X(\ ^{13} \CO)}{5 \times 10^{5}}  \\ &  \frac{1}{1-\exp(-10.58 \ \rm{K}/\rm{T_{ex}})} \exp{\left(\frac{5.29 \ \rm{K}}{\rm{T_{ex}}} \right)} \left[1+\left(\frac{\rm{T_{ex}}}{5.29  \ \rm{K}}\right)^2\right]^{1/2}
\end{split}
\end{equation}

Observations measure the brightness temperature of the lines, $^{12} \rm{T_B}$ and $^{13} \rm{T_B}$. If the $^{12}\CO$ \ (J=2-1) line is optically thick, we can find the excitation temperature from the measured brightness temperature using the relation :

\begin{equation}
\label{eqn:TB_12}
^{12} \rm{T_B} (v) = 11.066 \ \rm{K} \left(\frac{1}{\exp(11.06 \ \rm{K}/\rm{T_{ex}})-1}-0.0176\right)
\end{equation}

While solving the above equation we use the peak brightness temperature of the emission component.

With $\rm{T_{ex}}$ known for a particular emission feature we invert the following relation to find the optical depth of the $^{13}\CO$ \ (J=2-1) line in each channel.
\begin{equation}
\label{eqn:TB_13}
^{13} \rm{T_B} = 10.578 \rm{K} \left(\frac{1}{\exp(10.58 \  \rm{K}/\rm{T_{ex}})-1}-0.0212\right)(1-e^{-\tau_{13}(v)})
\end{equation}

We then integrate the  $^{13}\CO$ \ (J=2-1) optical depth map over the velocity range of the emission component and use equation \ref{eq:13_upper} to solve for $\NH$.


\bsp	
\label{lastpage}
\end{document}